\documentstyle[aps,prl,floats]{revtex}
\begin{document}
\renewcommand{\textfraction}{0.10}
\renewcommand{\topfraction}{1.0}
\renewcommand{\bottomfraction}{1.0}
\flushbottom

\twocolumn[
\title{Molecular geometry optimization with a genetic algorithm}
\author{D.M. Deaven and K.M.~Ho}
\address{Physics Department, Ames Laboratory USDOE, Ames, IA 50011}
\date{January 18, 1995 (accepted for publication in {\it Physical
Review Letters})}
\maketitle
\begin{abstract}
\widetext
\advance\leftskip by 57pt
\advance\rightskip by 57pt

We present a method for reliably
determining the lowest energy structure
of an atomic cluster in an arbitrary model potential.  The
method is based on a genetic algorithm, which operates on a
population of candidate structures to produce new candidates with
lower energies.  Our method dramatically outperforms simulated
annealing, which we demonstrate by applying the genetic algorithm
to a tight-binding model potential for carbon.  With this potential,
the algorithm efficiently finds fullerene cluster structures
up to ${\rm C}_{60}$ starting from random atomic coordinates.

\end{abstract}
]
\narrowtext

Advances in computer technology have made
molecular dynamics simulations
more and more popular in studying the behavior of
complex systems. Even with modern-day computers, however,
there are still two main limitations facing atomistic simulations:
system size and simulation time.  While recent developments in
parallel computer design and algorithms have made considerable
progress in enlarging the system size that can be accessed using
atomistic simulations, methods for shortening
the simulation time still remain relatively unexplored.

One example where such methods will be useful is in the determination
of the lowest energy configurations of a collection of atoms.
Because the number of candidate local energy minima grows exponentially with
the number of atoms, the computational effort scales exponentially with
problem size, making it a member
of the $NP$-hard problem class\cite{wille85}.  In practice, realistic
potentials describing covalently bonded materials
possess significantly more rugged energy landscapes
than the two-body potentials
addressed by the authors of Ref.\ \cite{wille85},
further increasing the difficulty.
Attempts to use simulated annealing to find the
global energy minimum in these systems are frustrated by
high energy barriers which trap the simulation
in one of the numerous metastable configurations.
Thus an algorithm is needed
which can `hop' from one minimum to another and permit an efficient
sampling of phase space.

In this letter, we will describe the application of such an
algorithm to the concrete example  of
determining the ground state structure of small atomic clusters.
The most interesting clusters are those which lie in the transition
range between molecules and bulk matter. These are precisely the
ones which can be expected to have unusual structures which are
unrelated to either the bulk or molecular limits.
For a few atoms, the ground state can sometimes be found by a brute
force search of configuration space.
For up to ten or twenty atoms, depending upon the potential,
simulated annealing may be employed to generate some
candidate ground state
configurations\cite{kirkpatrick83}.  For more
atoms than this, the simulation time required to find the
minimum by simulated annealing is usually prohibitive, because
evaluations
of the potential and forces are too expensive.
In this regime one is left with
judicious guessing of likely candidate ground state structures.

Our approach is based on the genetic algorithm (GA), an
optimization strategy inspired by the Darwinian evolution
process\cite{holland}.
Starting with a population of candidate structures,
we relax these candidates to the nearest local minimum. Using
the relaxed energies as the criteria of fitness, a
fraction of the population is selected as ``parents.''
The next generation of candidate structures is produced
by ``mating'' these parents.
The process is repeated until the ground state structure
is located.

We have applied this algorithm to optimize the geometry
of carbon clusters up to ${\rm C}_{60}$. In all cases we studied, the
algorithm efficiently finds the ground state structures starting from
an unbiased population of random atomic coordinates. This performance
is very impressive since carbon clusters are bound by strong directional
bonds which result in large energy barriers between different isomers.
Although
there have been many previous attempts to generate the C$_{60}$
buckyball structure from simulated annealing,
none has yielded the ground state structure\cite{attempt-c60}.

{\bf Method} --
Before presenting our results, we will describe
our genetic algorithm procedure in more detail.   The choice
of mating procedure is the central choice one must make in
constructing a genetic algorithm.  In an efficient algorithm, it
should impart important properties of the parent clusters
to the children. A common choice\cite{goldberg89}
is to first map the physical structure onto a binary number string,
then use string recombination as a mating
procedure. Such an approach has been applied to optimize
the packing structure of small molecular clusters
and the conformation of some molecules\cite{xiao93}.
We found that it is not very
efficient, however, when used to optimize the geometry of atomic
clusters.
This is because the mating operation does not preserve the
characteristics of the parents.

In the present work, we
represent an atomic cluster by the list of $N$
atomic cartesian coordinates ${\bf x}_i$ in arbitrary order,
\begin{equation}
{\cal G} = \{{\bf x}_1, {\bf x}_2, \ldots , {\bf x}_N\}.
\label{eq-ourrep}
\end{equation}
Our mating operator $P: P({\cal G}, {\cal G}') \rightarrow {\cal G}''$
performs the following action upon two
parent geometries ${\cal G}$ and ${\cal G}'$ to produce a child
${\cal G}''$. First, we choose a random plane passing
through the center of mass of each parent
cluster. We then cut the parent clusters
in this plane, and assemble the child ${\cal G}''$ from
the atoms of ${\cal G}$ which lie above the plane, and
the atoms of ${\cal G}'$ which lie below the plane.
If the child generated in this manner does not contain the correct
number of atoms, the parent clusters are
translated an equal distance in opposing
directions normal to the cut plane
so as to produce a child ${\cal G}''$ which contains the
correct number of atoms.

Relaxation to the nearest local minimum is
performed with conjugate-gradient minimization or molecular
dynamics quenching.
Typically, about 16 conjugate-gradient steps
or about 30 molecular dynamics steps are applied to a new geometry
before a decision is made whether further optimization is warranted.

We preferentially select parents with lower energy from
$\{{\cal G}\}$. The probability $p({\cal G})$ of
an individual candidate ${\cal G}$ to be selected for mating is
given by the Boltzmann distribution
\begin{equation}
p({\cal G}) \propto \exp(-E({\cal G}) / T_{\rm m}),
\label{prob-eq}
\end{equation}
where $E({\cal G})$ is the energy per atom of the candidate ${\cal G}$, and
the mating `temperature' $T_{\rm m}$ is chosen to be roughly equal to
the range of energies in $\{{\cal G}\}$.

In some cases, described in the next section, we found it necessary to
apply mutations to members of the population.
We define a mutation operator $M: M({\cal G}) \rightarrow {\cal G}'$
which performs one of two functions with equal probability.
The first mutation function
moves the atoms in ${\cal G}$ a random distance (of the same order
as a bond length), in a random direction, a random
number of times (between 5 and 50),
while separating unphysically close atoms between each step.
The second mutation function implements a simple
search for an adjacent watershed in the
potential energy hypersurface.  We employ an
algorithm\cite{longpaper}
which takes a random number of steps
in atomic coordinate space.
At each step the algorithm changes direction so as to
maintain travel along a direction slightly uphill to
an equipotential line.  The result of this is generally a
high-energy cluster, but one which lies in an adjacent
watershed region of $E(\{{\bf x}\})$.

We maintain a population $\{{\cal G}\}$ of $p$ candidates, and
create subsequent generations as follows.
Parents are continuously chosen from $\{{\cal G}\}$ with probability
given by Eq.~(\ref{prob-eq}) and mated using the mating procedure
described above. A fraction $\mu$ of the
children generated in this way are mutated; $\mu=0$ means no mutation
occurs.  The (possibly mutated) child is relaxed to the nearest local
minimum and selected for inclusion in the population if its energy
is lower than another candidate in $\{{\cal G}\}$.

This procedure requires the algorithm to
keep track of a large number of candidates in
$\{{\cal G}\}$, since the population generally
becomes filled with almost
identical low-energy candidates. These duplicated efforts reduce the
algorithm's efficiency. To prevent this,
we introduce an energy resolution $\delta E$,
and allow new entries to $\{{\cal G}\}$ only if there are no other
candidates already in $\{{\cal G}\}$ whose energy is within $\delta E$
of the new entry's energy.

\begin{figure}[t]
\setlength{\unitlength}{1 mm}%
\begin{picture}(80,75)
\put(18,13){\includegraphics{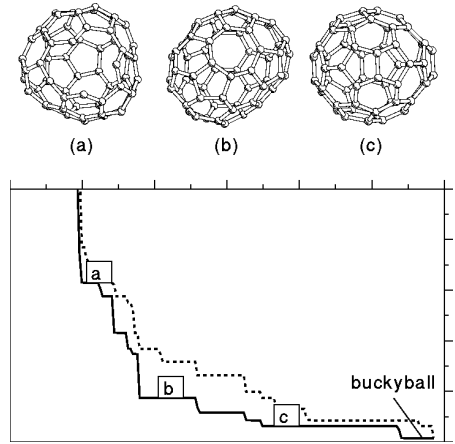}}
\end{picture}
\caption[c60-fig-cap]{
Generation of the ${\rm C}_{60}$ molecule, starting from random
coordinates, using the genetic algorithm described by the text with
4 candidates ($p=4$) and no mutation ($\mu = 0$).
The energy per atom is
plotted for the lowest energy (solid line) and highest energy (dashed
line) candidate structure in $\{{\cal G}\}$ as a function of the number of
genetic mating operations $P$ (see text) that have been applied.
Several of the intermediate structures which contain defects
are illustrated at top: (a) contains
one 12-membered ring and two 7-membered rings, (b) contains a 7-membered
ring, (c) contains the correct distribution of pentagons and
hexagons, but two pentagons are adjacent. The ideal
icosahedral buckyball structure
is achieved shortly after 5000 genetic operations.}
\label{c60-fig}
\end{figure}

{\bf Results} --
To illustrate the method, we use a tight-binding model
for carbon, described elsewhere\cite{potentialref}. This potential
accurately describes the energetics of fullerene structures.

Fig.\ \ref{c60-fig} shows the model potential energy of the lowest
and highest energy ${\rm C}_{60}$ cluster in $\{{\cal G}\}$
versus the number of genetic mating
operations performed with no mutation ($\mu=0$)
on a population of $p=4$ candidates,
starting from coordinates chosen at random.
We used a mating temperature $T_{\rm m} = 0.2$~eV/atom,
and an energy resolution $\delta E = 0.01$~eV/atom.
This cluster is too large for unbiased simulated
annealing\cite{bernholc86} to arrive
at the correct global minimum (the icosahedral buckminsterfullerene
cage).  As Fig.\ \ref{c60-fig} illustrates, the
genetic algorithm correctly generates the cage after roughly
5000 mating operations.

Fig.\ \ref{c60-fig} illustrates several generic features of the
algorithm.  During the initial few generations,
the energy drops very quickly and the population soon consists of
reasonable candidates, similar to what would be observed with
simulated annealing.  This initial period is usually a small fraction
of the total time spent by the algorithm.
The rest of the time
is spent in an end game, where the remaining
defects in the structure are removed
(Fig.\ \ref{c60-fig} (a) -- (c)).
The general behavior of the genetic algorithm is remarkably
resistant to changes in the details of the algorithm.  The
${\rm C}_{60}$ cage is found reliably over a wide range of values of the
mating temperature $T_{\rm m}$, number of candidates $p$, and the
number of conjugate-gradient optimizations performed upon each
application of $O$.  In addition, the use of schemes other than
Eq.~(\ref{prob-eq}) for selecting parents from $\{{\cal G}\}$ also
leads to the correct final answer.  For example, we tried using
equal mating probabilities $p({\cal G})$
for all candidates regardless of
energy, as well as a probability linear in the energy.  All of these
variations produced genetic algorithms which worked satisfactorily.

In cases with several competing low energy states, it
is sometimes advantageous to investigate the minimization of a number
of ``ecologies,'' that is, to repeat
the above process with different starting populations.
For example, in smaller clusters of carbon atoms, a bimodal mass
spectrum has been observed in laser vaporization
experiments\cite{rohlfing84}, and this
has been interpreted\cite{tomanek91} as evidence that two regimes of
${\rm C}_N$ cluster growth exist:
for $N \alt 25$, mono- and polycyclic rings are formed, while for
$N \agt 25$, fullerene cages are formed. Thus, for clusters around this
size, there is a competition between cage-like, ring-like and cap-like
structures.  Searches for the global energy minimum must surmount the
difficulty of becoming trapped in one of these
structural classes.

\begin{figure}[t]
\setlength{\unitlength}{1 mm}%
\begin{picture}(80,116)
\put(18,13){\includegraphics{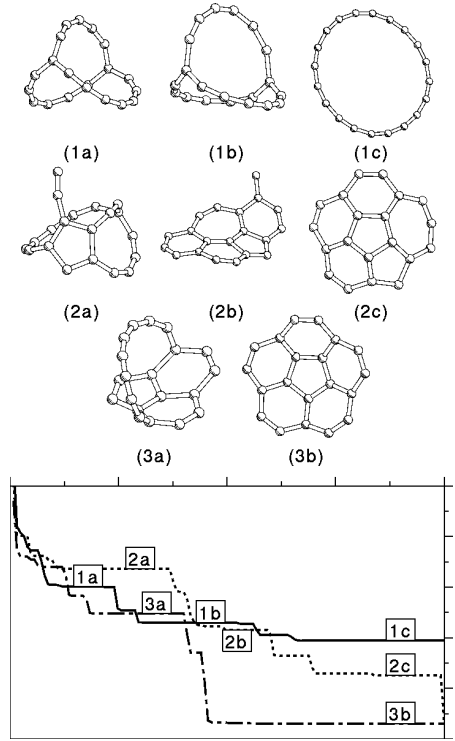}}
\end{picture}
\caption[c20-fig-cap]{
Running the genetic algorithm on ${\rm C}_{20}$.
The solid line shows the generic
lowest energy structure when the algorithm is
run with no mutation ($\mu=0$); the structures (1a) - (1c) are present
in the population at the times indicated.  Essentially all ecologies
get trapped in monocyclic rings (1c).  The dashed line (structures
(2a) - (2c)) and the dot-dashed line (structures (3a) and (3b)) illustrate
the results when mutation is added ($\mu=0.05$).}
\label{c20-fig}
\end{figure}

\begin{figure}[t]
\setlength{\unitlength}{1 mm}%
\begin{picture}(80,93)
\put(18,13){\includegraphics{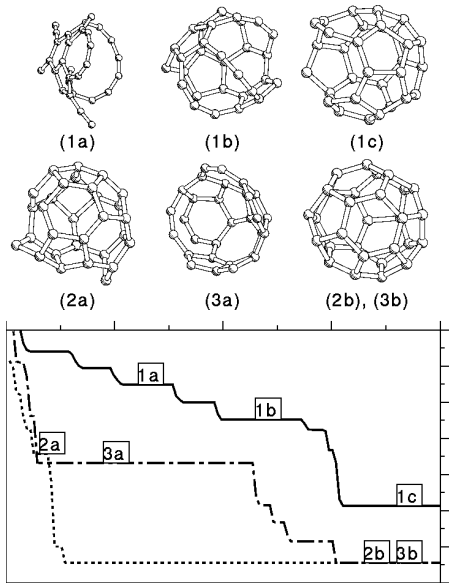}}
\end{picture}
\caption[c30-fig-cap]{
Running the genetic algorithm on ${\rm C}_{30}$.
The solid line shows the lowest energy structure when the algorithm is
run with no mutation ($\mu=0$) for an ecology that failed to find the
minimum energy configuration (a fullerene cage) within 4000
genetic operations.  The structures (1a) - (1c) are present
in the population at the times indicated.  The structure (1c) resulting
after 4000 genetic operations is a cage, and is eventually reduced to
the perfect fullerene cage even with $\mu=0$.
The broken lines illustrate two
$\mu=0.05$ ecologies which arrive at the perfect cage (2b) via
distinct routes (2a), (3a).}
\label{c30-fig}
\end{figure}

Figs.\ \ref{c20-fig} and\ \ref{c30-fig}
show the results of running the genetic
algorithm on ${\rm C}_{20}$ and ${\rm C}_{30}$ clusters, using the
same parameters $p=4$ and $T_{\rm m}=0.2$~eV/atom that were used to
generate the ${\rm C}_{60}$
cage.  The solid line in Fig.\ \ref{c20-fig} illustrates the generic
result for ${\rm C}_{20}$ when no mutation is used ($\mu=0$).
The lowest energy
structure for ${\rm C}_{20}$ in the model potential is a polycyclic
cap with energy $-8.671$ eV/atom, and the fullerene
cage structure is not far above, with energy
$-8.613$ eV/atom. Nevertheless, only a small fraction
of the $\mu=0$ genetic algorithm ecologies
find one of these structures within 4000 genetic operations.
Instead, the ecologies get `trapped' in monocyclic rings with energy
$-8.503$ eV/atom (Fig.\ \ref{c20-fig} (1c)).
The cap and the cage structures can be found
for ${\rm C}_{20}$, however, if we include mutations in our algorithm or,
equivalently,
by using molecular dynamics annealing for the relaxation process.
For example, with $\mu=0.05$, about 25\% of the ecologies
find the polycyclic cap (Fig.\ \ref{c20-fig}, broken lines).

In the case of ${\rm C}_{30}$, the lowest energy structure
in the model potential is a
fullerene cage, and roughly 80\% of the $\mu=0$ ecologies find it within
4000 genetic operations. The remaining 20\% form cages, but not
quickly enough to find the fullerene (Fig.\ \ref{c30-fig}, solid
line).
With mutations, convergence to the fullerene cage is
greatly increased. Essentially all
of the $\mu=0.05$ ecologies find the ${\rm C}_{30}$
cage within 4000 genetic operations
(Fig.\ \ref{c30-fig}, broken lines).
The role of mutation in the algorithm is to allow searches for
alternate structural classes. Referring to Fig.\ \ref{c20-fig},
one sees precipitous drops in
energy when a new class of candidate is discovered.  In the case of
${\rm C}_{30}$, the cage structural class appears even with $\mu=0$
but is more efficiently reduced to the perfect structure
when $\mu\not=0$.

We emphasize that mutation by itself does not efficiently lower the
energy of a population.  We found that application of the mutation
operator $M$ in the absence of mating leads to a drastic
decrease in the efficiency of the optimization process.

{\bf Discussion} --
Like simulated annealing, the genetic algorithm requires repeated
evaluation of the energy and forces within the model potential.
The higher efficiency of the genetic algorithm, however, allows
convergence to low-energy candidates in larger clusters than is possible
with simulated annealing. We are currently applying the method to
larger carbon clusters and will present those results
elsewhere\cite{longpaper}.
In addition, we have applied the algorithm
to systems other than carbon clusters, and our preliminary
findings indicate that the algorithm is efficient
over a broad class of structural optimization problems.
For example, we have successfully applied the method to
bulk and surface geometries, with a suitably modified
mating operator $P$.

The efficiency of the present algorithm may
be increased in special cases when the class of desired structures is
assumed, and a more complicated
mapping between the genetic representation (genotype) and the cluster
structure (phenotype) could be employed. For
instance, in the case of the larger carbon fullerene clusters we
expect that a representation in terms of a face-dual
model\cite{face-dual} would lead to
rapid convergence, since only cage structures would be investigated.


While the artificial dynamics of the genetic algorithm cannot be
expected to reproduce the natural annealing process in
which atomic clusters are
formed, we found that the intermediate structures located by the genetic
algorithm on its way to the ground state structure are very similar to
the results of simulated annealing. Thus it appears that
the same kinetic factors which influence
the annealing process also affect the ease with which a particular
candidate is generated by the genetic algorithm.
If this is true,
the genetic algorithm results presented here can be viewed
as analogous to those of greatly extended conventional
simulated annealing runs.
More work needs to be done to determine if this is indeed the case.

{\bf Acknowledgments} --
We thank C.Z.~Wang for useful discussions, and
J.R.~Morris and J.L.~Corkill for useful comments on the manuscript.
Part of this work was made possible by the Scalable Computing
Laboratory, which is funded by the Iowa State University and Ames
Laboratory.
Ames Laboratory is operated for the U.S.~Department of Energy by Iowa
State University under contract no. W-7405-Eng-82.  This work was
supported by the Director for Energy Research, Office of Basic Energy
Sciences, and the High Performance Computing and Communications
initiative.

\end{document}